\documentclass{article}

\usepackage{ijcai05}
\usepackage{times}
\usepackage{graphicx}
\usepackage{amssymb,amsmath}
\usepackage{algorithmic}
\usepackage{algorithm}
\usepackage{latexsym}

\title{An n-ary Constraint for the Stable Marriage Problem\thanks{The first author is supported by EPSRC. Software support was given by an ILOG SA's academic grant.}}

\author{Chris Unsworth, Patrick Prosser \\ Department of Computing Science \\
  University of Glasgow, Scotland \\ \{chrisu,pat\}@dcs.gla.ac.uk}

\begin{document}

\maketitle

\begin{abstract} 
We present an n-ary constraint for the stable marriage problem. This constraint acts
between two sets of integer variables where the domains of those variables represent preferences.
Our constraint enforces stability and disallows bigamy. For a stable marriage instance with $n$ men and $n$ women
we require only one of these constraints, and the complexity of enforcing arc-consistency is $O(n^2)$ which 
is optimal in the size of input. Our computational studies show that our n-ary constraint is significantly faster
and more space efficient than the encodings presented in \cite{cp01}. We also introduce a new problem to the 
constraint community, the sex-equal stable marriage problem.
\end{abstract}
\section{Introduction}
In the Stable Marriage problem (SM) \cite{gs62,gi89} we have $n$ men and $n$ women. Each man ranks the $n$ women 
into a preference list, as do the women. The problem is then to produce a matching of men to women such that 
it is stable. By a matching we mean that there is a bijection from men to women, and by stable we
mean that there is no incentive for partners to divorce and elope. A matching is unstable if there are 
two couples $(m_i,w_j)$ and $(m_k,w_l)$ such that $m_i$ prefers $w_l$ to his current partner $w_j$, and 
$w_l$ prefers $m_i$ to her current partner $m_k$. 

Figure 1 is an instance of the stable marriage problem, and has 6 men and 6 women. Figure 1 shows the problem 
initially, with each man and woman's preference list. Figure 2 shows the intersection of the male and 
female-oriented Gale-Shapley lists (GS-lists) \cite{gi89}, where the GS-lists are reduced preference lists. A 
man-optimal (woman-pessimal) stable matching can now be found by  marrying men (women) to their most (least)
preferred choices in there GS-lists. Conversely, we can produce a woman-optimal (man-pessimal) 
matching by marrying women (men) to their most (least) preferred choice in their GS-lists. An instance of
SM admits at least one stable matching and this can be found via the Extended Gale-Shapley algorithm in 
time $O(n^2)$, where there are $n$ men and $n$ women.

\begin{figure}[htb]
\begin{center}
\begin{tabular}{||c||c||} \hline
Men's lists & Women's lists \\ \hline
1: 1 3 6 2 4 5 & 1: 1 5 6 3 2 4 \\
2: 4 6 1 2 5 3 & 2: 2 4 6 1 3 5 \\
3: 1 4 5 3 6 2 & 3: 4 3 6 2 5 1 \\
4: 6 5 3 4 2 1 & 4: 1 3 5 4 2 6 \\
5: 2 3 1 4 5 6 & 5: 3 2 6 1 4 5 \\
6: 3 1 2 6 5 4 & 6: 5 1 3 6 4 2 \\ \hline
\end{tabular}
\end{center}
\caption{An SM instance with 6 men and 6 women}
\begin{center}
\begin{tabular}{||l||l||} \hline
Men's lists & Women's lists  \\ \hline
1: 1     & 1: 1 \\
2: 2     & 2: 2 \\
3: 4     & 3: 4 6 \\
4: 6 5 3 & 4: 3 \\
5: 5 6   & 5: 6 4 5 \\
6: 3 6 5   & 6: 5 6 4 \\ \hline
\end{tabular}
\end{center}
\caption{the corresponding GS-lists}
\label{SMinstance}
\end{figure}

We present a simple constraint encoding for the stable marriage problem. We introduce a specialised 
n-ary constraint with only three methods, where each method is no more than six lines of code. We show how 
enforcing arc-consistency in this encoding results in the male-oriented Gale-Shapley lists. This minimal
encoding cannot be used in search and only achieves directed arc-consistency, from men to women.
We then go on to 
show how we can extend this encoding by introducing a modest amount of additional code, such that the encoding 
can be used in search, can be embedded in richer impure problems where the stability of marriages is only part 
of a larger problem, 
and the male and female oriented GS-lists are produced. Our empirical results suggest, that although our encodings 
has $O(n^2)$ time complexity, the same as the optimal encoding proposed in \cite{cp01}, our constraint significantly 
outperforms this encoding in both space and time. 

\section{The Extended Gale-Shapley Algorithm (EGS)} 
We now describe the male-oriented Extended Gale-Shapley (EGS) algorithm (shown in Figure \ref{EGSalg}).
In particular, we explain what is meant by a {\em proposal}, an {\em engagement}, and for a man to become {\em free}.
We will use this later to show that this algorithm and our constraint encoding are equivalent.

The EGS algorithm \cite{gi89} produces a stable matching between men $m_1$ to $m_n$ and
women $w_1$ to $w_n$, where each man (woman) ranks each of the women (men) into preference order. Via a process
of proposals from  men to women the algorithm delivers reduced preference lists, called GS-lists
(Gale-Shapley lists), such that if each man (woman) is paired with his (her) best (worst) partner in their
GS-list the marriages will be stable.\footnote{Strictly speaking, the given algorithm  produces MGS-lists, the
male GS-lists. But for the sake of brevity we will refer to them as GS-lists.}

\begin{figure}[h]
\begin{small}
\begin{verbatim}
 1  assign each person to be free
 2  WHILE (some man m is free)
 3  DO BEGIN
 4     w := first woman on m's list
 5     IF (some man p is engaged to w)
 6     THEN assign p to be free
 7     assign m and w to be engaged
 8     FOR (each successor p of m on w's list)
 9     DO BEGIN
10        delete p from w's list
11        delete w from p's list
12        END
13     END
\end{verbatim}
\end{small}
\caption{The male-oriented Extended Gale/Shapley algorithm.}
\label{EGSalg}
\end{figure}

We will assume that we have an instance $I$ of the stable marriage problem, and that for any person $q$ in $I$, 
$PL(q)$ is the ordered
list of persons in the original preference list of $q$ and $GS(q)$ is the ordered list of people in the
GS-list for $q$, and initially $GS(q)$ equals $PL(q)$. In a {\em proposal} from man $m$ to
woman $w$, $w$ will be at the head of the man's GS-list $GS(m)$. This leads to an {\em engagement} where 
$m$ is no longer free and all men that $w$ prefers less than $m$ are removed from her GS-list, i.e. 
the last entry in $GS(w)$ becomes $m$. 
Further, when a man $p$ is removed from $GS(w)$ that woman is also removed from his GS-list, i.e.
$w$ is removed from $GS(p)$, consequently bigamy is disallowed. Therefore $m$ and $w$ are engaged when 
$m$ is no longer free, $w$ is head of $GS(m)$, and $m$ is at the tail of $GS(w)$. A man $p$ becomes {\em free} when $p$
was engaged to $w$ (i.e. the head of $GS(p)$ is $w$) and $w$ receives a proposal from man $m$ that she prefers
to $p$. On becoming free, $p$ is added to the list of free men and $w$ is removed from $GS(p)$.

The algorithm starts with all men free and placed on a list (line 1). The algorithm
then performs a sequence of proposals (lines 2 to 13). A man $m$ is selected from the free list (line 2),
and his most preferred woman $w$ is selected (line 4). If $w$ is engaged, then her partner $p$ becomes free.
The pair $m$ and $w$ then become engaged (lines 7 to 12).

\section{Preliminaries}
We assume that the men and women's preference lists have been read into two 2-dimensional integer arrays $mpl$ 
and $wpl$ respectively. $mpl[i]$ is the preference list for the $i^{th}$ man where $mpl[i][j]$ is the i$^{th}$ 
man's j$^{th}$ preference, and similarly $wpl[j]$ is the preference list for the $j^{th}$ woman. Using our 
problem in Figure 1, if we consider our $3^d$ man he will have a preference list $mpl[3] = (1,4,5,3,6,2)$.

We also assume we have the inverse of the preference lists, i.e. $mPw$ and $wPm$, where
$mPw[i][j]$ is the $i^{th}$ man's preference for the $j^{th}$ woman and $wPm[k][l]$ 
is the k$^{th}$ woman's preference for the l$^{th}$ man.
Again, considering the 3$^d$ man in Figure 1, his inverse preference list will be $mPw[3] = (1,6,4,2,3,5)$,
$mPw[3][2]$ is his preference for the $2^{nd}$ woman, and that is
6, i.e. woman 2 is in the $6^{th}$ position of man 3's preference list.\footnote{The inverse of the preference lists
can be created when reading in the preference lists such that $mPw[i][mpl[i][j]] = j$, and this does not affect
the overall complexity of constructing our model.}

We associate a constrained integer variable with each man and each woman, such that $x[i]$ is 
a constrained integer variable representing the $i^{th}$ man $m_{i}$ in stable marriage instance $I$ 
and has a domain $dom(x[i])$ initially of 1 to $n$. Similarly, we have an array of constrained integer variables 
for women, such that $y[j]$ represents the $j^{th}$ woman $w_j$ in $I$. The values in the domain of a variable 
correspond to preferences, such that if variable $x[i]$ is assigned the value $a$ this corresponds to $m_i$ being 
married to his $a^{th}$ choice of woman, and this will be woman $mpl[i][a]$. For example, if $x[2]$ (in Figure 1)
is set to 3 then this corresponds to $m_2$ marrying his $3^d$ choice, $w_1$ (and conversely 
$y[1]$ would then have to be assigned the value 5). Again referring to Figure 1 our $6^{th}$ man's
domain is $dom(x[6]) = (1,2,3,4,5,6)$, as is everyone else's, and in Figure 2 $dom(x[6]) = (1,4,5)$. 
We also assume that we have the following functions, each being of $O(1)$ complexity, that operate over 
constrained integer variables:

\begin{itemize}
\item
$getMin(v)$ delivers the smallest value in $dom(v)$.
\item
$getMax(v)$ delivers the largest value in $dom(v)$.
\item
$getVal(v)$ delivers the instantiated value of $v$.
\item
$setMax(v,a)$ sets the maximum value in $dom(v)$ to be min$(getMax(v),a)$.
\item
$setVal(v,a)$ instantiates the variable $v$ to the value $a$.
\item
$remVal(v,a)$ removes the value $a$ from $dom(v)$.
\end{itemize}

We assume that constraints are processed by an arc-consistency algorithm such as 
AC5 \cite{ac5} or AC3 \cite{ac3}. That is, the algorithm has a stack of constraints that are awaiting revision and
if a variable loses values then all the constraints that the variable is involved in are added to the stack
along with the method that must be applied to those constraints, i.e. the stack contains methods and their arguments.
Furthermore, we also assume that a call to a method, with its arguments, is only added to the stack if it is 
not already on the stack. We'll refer to this stack as the {\em call stack}.

\section{An n-ary Stable Marriage Constraint (SM2N)} 
We now give a description of our n-ary stable marriage constraint, where arc-consistency on such an encoding is 
equivalent to an application of the male-oriented EGS algorithm.
Note that the constraint as described minimally 
cannot be used within a search process, however we will later show how this can be done. Our constraint is n-ary in 
that it constrains $n$ men and $n$ women such that stability is maintained and bigamy is disallowed, although 
it achieves only 2-consistency.\footnote{A detailed explanation of just what we mean by 
2-consistency in this model is given in section 6.}
In a stable marriage 
problem with $n$ men and $n$ women we will then require only one of these constraints. We now start by describing 
the attributes of the constraint and the three methods that act upon it.
We will use a java-like pseudo-code such that the $.$ (dot) operator is an attribute selector, such 
that $a.b$ delivers 
the $b$ attribute of $a$.

\subsection{The attributes}
A n-ary stable marriage constraint (SM2N) is an object that acts between $n$ men and $n$ women, and has the following 
attributes:
\begin{itemize}
\item
$x$ and $y$ are constrained integer variable arrays representing the men and women that are constrained, such that 
$x[i]$ is the constrained integer variable corresponding to $m_i$ and $y[j]$ corresponds to $w_j$.
\item
$xpl$ and $ypl$ are 2-dimensional integer arrays which contain the male and female preference lists respectively,
such 
that $xpl[i]$ equals $PL(m_i)$ and $xpl[i][j]$ contains $m_i$'s $j^{th}$ choice woman.
\item
$xPy$ and $yPx$ are 2-dimensional integer arrays which contain the male and female inverse preference 
lists respectively, 
such that $xPy[i][j]$ contains man $i$'s preference for $w_j$.
\item
$yub$ is an array of integer variables which contain the previous upper bounds of all $y$ variables. 
All are set to $n$ at
the start of search and are updated by the deltaMax(i) method detailed below.
\end{itemize}

\subsection{The propagation methods}
We now describe three methods that achieve male-oriented arc-consistency.

\subsubsection{deltaMin(i)}
This method is called when the lower bound of $dom(x[i])$ increases. The lower bound of $dom(x[i])$ increasing
signifies that $m_i$ has been rejected by his favourite choice of partner and thus must propose to his new favourite 
available partner. To do this we first find $m_i$'s favourite available partner $w_j$ (line 2), then remove 
all men from 
the list of $w_j$ she likes less than $m_i$ (line 3).

\begin{small}
\begin{verbatim}
  1.   deltaMin(i)
  2.    j = xPy[i][getMin(x[i])]
  3.    setMax(y[j],yPx[j][i])
\end{verbatim}
\end{small}

\subsubsection{deltaMax(j)}
This method is called when the upper bound of $dom(y[j])$ is reduced. To maintain consistency $w_j$ needs to 
be removed from the domains of all men that have been removed from her domain. This is done by looping 
once for each value that has been removed from the tail of $dom(y[j])$ since the last call to deltaMax(j) (line 2). 
Within the loop a $m_i$ that has been removed from $dom(y[j])$ is selected (line 3) and then $w_j$ is removed 
from $dom(x[i])$. When all relevant men have had their domains' altered (line 5) $yub$ is updated (line 6).

\begin{small}
\begin{verbatim}
  1.   deltaMax(j)
  2.    FOR (k = getMax(y[j])+1 to yub[j])
  3.     i = yPx[j][k]
  4.     remVal(x[i],xPy[i][j])
  5.    END FOR LOOP
  6.    yub[j] = getMax(y[j])
\end{verbatim}
\end{small}

\subsubsection{init()}
The $init$ method is called when the constraint is created, and is simply a call to $deltaMin$ for each 
of the $n$ men variables.
\begin{small}
\begin{verbatim}
  1.   init()
  2.    FOR (i = 1 to n)
  3.     deltaMin(i) 
  4.    END FOR LOOP

\end{verbatim}
\end{small}

\section{Comparison to EGS}
We now compare the behaviour of our n-ary constraint model (SM2N) to the male-oriented EGS 
algorithm. In our comparison we will describe steps in the EGS algorithm in {\em italics} 
and the SM2N constraint encoding in normal font. Sometimes we will use $m$ and $w$
as a particular person (rather than $m_i$ and $w_j$), and $x$ and $y$ as particular variables (rather than 
$x[i]$ and $y[j]$) for sake of brevity. Additionally, we assume we have
the function $fiance(y[i])$ and that it delivers the integer $k$ where 
$k = wpl[i][max(dom(y[i])]$, i.e. $x[k]$ is the least preferred partner of $y[i]$.

\begin{itemize}
\item
{\em Initially the EGS algorithm sets all men to be free by adding them to the free list (line 1).}
Equivalently, when propagation starts the call to $init()$ will cause the set of calls 
$\{deltaMin(i) | 1 \leq i \leq n\}$ to be added to the empty call stack.
\item
{\em EGS picks a man $m$ from the free list and he then proposes to his first choice woman $w$ (lines 4 to 7).}  
Initially the call stack will contain $n$ calls to the $deltaMin$ method, called directly via $init$.
When executing the call $deltaMin(i)$, man $x[i]$ will make the equivalent of a proposal to his first choice woman 
(as described next).
\item
{\em When $m$ makes a proposal to $w$ all values that appear in $GS(w)$ after the proposing man are 
removed (lines 8 to 10), i.e. they become engaged.} 
When the call $deltaMin(i)$ is made, where $y[j]$ is $x[i]'s$ favourite, the maximum of  $dom(y[j])$ is set to $y[j]'s$
preference for $x[i]$, therefore removing all less preferred men. Effectively, $x[i]$ and $y[j]$ become engaged.
\item
{\em To maintain monogamy EGS removes the newly engaged woman from the GS-lists of all men that 
have just been removed from her preference list (line 11).}
From the action above, the maximum of $dom(y[j])$ has been lowered, consequently a call to $deltaMax(j)$ will be 
added to the call stack. In that call to $deltaMax(j)$, $y[j]$ is removed from $dom(x[k])$ for all $k$ where 
$k$ has been removed from the tail of $dom(y[j])$. Therefore, $x[k]$ and $y[j]$ can never be married. 
\item
{\em In EGS, if $m$ makes a proposal to $w$, who is already engaged to $p$, then $w's$
previous fiance $p$ is assigned to be free and added to the free list (lines 5 and 6.)} On initiating the call
$deltaMin(i)$ where $y[j]$ is $x[i]'s$ favourite available woman, $y[j]'s$ fiance corresponds to the maximum value in 
$dom(y[j])$, because all less preferred men will have been removed (as above).  Therefore if $y[j]$ receives a 
proposal from $x[i]$ via the call $deltaMin(i)$, and $y[j]$ prefers $x[i]$ to her current fiance $x[k]$ (where
$k = fiance(y)$) 
the maximum of $dom(y[j])$ will be set lower than her preference for $x[k]$ and therefore her preference for 
$x[k]$ will 
be removed from $dom(y[j])$. Consequently, the call $deltaMax(j)$ will then be put on the call stack, which will remove
$x[k]'s $ preference for $y[j]$ from $dom(x[k])$. Because $y[j]$ was $x[k]'s$ previous favourite, $x[k]'s$ 
preference for $y[j]$ would 
have been $min(dom(x[k]))$. Therefore removing that value will increase $x[k]'s$ domain minimum, 
and the call $deltaMin(k)$ will then be added to the stack. And this effectively assigns man $x[k]$ to be free.
\end{itemize}

\section{Arc-consistency in the Model}
On the completion of arc-consistency processing, the variable domains can be considered as $GS-domains$. That is, 
$a \in dom(x[i]) \leftrightarrow w_{j} \in GS(m_{i}) \wedge j = mpl[i][a]$. Furthermore, 
$b \in dom(y[j]) \leftrightarrow m_{i} \in GS(w_{j}) \wedge i = wpl[j][b]$.

The GS-domains are 2-consistent such that if man $m_i$ is married to a woman $w_j$ 
(i.e. $x[i] = a \wedge a \in dom(x[i]) \wedge j = mpl[i][a]$) then any woman $w_l$ can then marry
some man $m_k$ without forming a blocking pair or a bigamous relationship. That is, for an arbitrary woman
$w_l$ there exists a value $b \in dom(y[l])$ such that 
$k = wpl[l][b] \wedge (mPw[i][j] < mPw[i][l] \vee wPm[l][k] < wPm[l][i]) \wedge i \neq k \wedge j \neq l$.
Furthermore if a man $m_i$ is married to a woman $w_j$ then any other man $m_k$ can then marry some woman $w_l$,
where $l \neq j$.

It is important to note, that although our constraint is n-ary it only achieves 2-consistency. It is our opinion
that the cost of achieving a higher level of consistency would be of little advantage. This is so because 
by maintaining 2-consistency, and using a suitable value ordering heuristic in the model during search we 
are guaranteed failure-free enumeration of all solutions \cite{cp01}.

In \cite{gi89} Theorem 1.2.2 it is proved that all possible executions of
the Gale-Shapley algorithm (with men as proposers) yield the same stable matchings. Our encoding 
mimics the EGS algorithm (as shown in section 5) and we claim (without proof) that the encoding
reaches the same fixed point for all ordering of the revision methods on the call stack.

\section{Complexity of the model}
In \cite{gi89} section 1.2.3 it is shown in the worst case there is at most $n(n-1)+1$ proposals that can be made by the 
EGS algorithm, and that the complexity is then $O(n^{2})$. We argue that the complexity of our SM2N encoding 
is also $O(n^{2})$. 
First we claim that the call to our method $deltaMin()$ is of complexity $O(1)$. The $deltaMax()$ method is of complexity 
$O(r)$, where $r$ is the number of values removed from the tail of variable since the last call to $deltaMax()$ for this
variable.

Because there are $n$ values in the domain of variable $y$ the worse case complexity for all possible calls to 
$deltaMax(j)$ is $O(n)$. Equally there are $n$ values in the domain of variable $x$ and thus the worse case 
complexity for all possible calls to $deltaMin(i)$ is $O(n)$. Therefore because there are $n$ $y$ variables and
$n$ $x$ variables, the total worst case complexity for all possible calls to $deltaMin(i)$ and $deltaMax(j)$ is $O(n^2)$.

\section{Enhancing the model}
The full GS-Lists are the union of the male and female Gale-Shapley lists remaining after executing
male and female oriented versions of EGS.  It has been proven that the same lists can be produced
by running the female orientated version of EGS on the male-oriented GS-lists \cite{gi89}. Because SM2N 
produces the same results as EGS the full GS-Lists can be produced in the same way.  But because of the 
structure of this specialised constraint it is also possible to combine the male and female orientated 
versions of SM2N into one constraint.  This combined gender free version of SM2N will then produce the 
full GS-List with only one run of the arc-consistency algorithm. To create the gender free version 
all of the methods presented in this paper
must then be symmetrically implemented from the male and female orientations.

The SM2N constraint as presented so far has only considered domain values being removed by the constraint's 
own methods.  If we were to use the constraint to find all possible stable matchings, unless arc consistency 
reduces all variable domains to a singleton, it will be necessary to assign and remove values from variable 
domains as part of a search process.  Therefore, we need to add code to SM2N to maintain consistency and 
stability in the event that domain values are removed by methods other than those within SM2N.  It is important 
to note that these external domain reductions could also be caused by side constraints as well as a search process.

There are four types of domain reduction that external events could cause: 
a variable is instantiated; a variable's minimum domain value is increased; a 
variable's maximum domain value is reduced; one or more values are removed from the interior of a
variable's domain. We now describe two additional methods, $inst$ and $removeValue$, and the enhancements required for
$deltaMin$. We note that $deltaMax$ does not need to change, and describe 
the required enhancements for incomplete preference lists.

\subsubsection{inst(i)}
The method $inst(i)$ is called when a variable $x[i]$ is instantiated.
\begin{small}
\begin{verbatim}
  1.   inst(i) 
  2.    For (k = 0 to getVal(x[i])-1)
  3.     j = xPy[i][k]
  4.     setMax(y[j],yPx[j][i]-1)
  5.    END FOR LOOP
  6.    j = xPy[i][getVal(x[i])]
  7.    setVal(y[j],yPx[j][i]) 
  8.    For (k = getVal(x[i])+1 to n)
  9.     j = xPy[i][k]
 10.     remVal(y[j],yPx[j][i]) 
 11.    END FOR LOOP
\end{verbatim}
\end{small}

This method removes all values from the set of $y$ variables to prevent variable $x[i]$ being involved in a blocking
pair or inconsistency. To prevent $x[i]$ from creating a blocking pair, all the values that corresponds to men
less preferred than $x[i]$, are removed from the domains of all women that $x[i]$ prefers to his assigned partner 
(lines 2-5). Since $x[i]$ is matched to $y[j]$, $y[j]$ must now
be matched to $x[i]$ (lines 6,7). To maintain consistency $x[i]$ is removed from the domains of all other women 
(lines 8-11)). The complexity 
of this method is $O(n)$ and because there are $n$ $x$ variables and each can only be instantiated once during 
propagation, the total time complexity of all possible calls to $inst(i)$ is $O(n^2)$.

\subsubsection{removeValue(i,a)}
This method is called when the integer value $a$ is removed from $dom(x[i])$, and this value
is neither the largest nor smallest in $dom(x[i])$.
\begin{small}
\begin{verbatim}
  1.   removeValue(i,a)
  2.    j = xPy[i][a]
  3.    remVal(y[j],yPx[j][i])
\end{verbatim}
\end{small}
The woman the value $a$ corresponds to is found (line 2) then $x[i]$ is removed from her domain (line 3), and this 
must be done to prevent bigamy.

\subsubsection{Enhancements to deltaMin(i)}
Up till now we have assumed that all values removed from the head of $dom(x[i])$ are as a result of $m_i$ being
rejected by some $w_j$. We now drop this assumption in the following enhanced version. In this method we add a
new variable array named $xlb$, and this is similar to the $yub$ array except it holds the previous lower bound of 
$x$. All elements in $xlb$ are initialised to $1$ and are updated and used only by the $deltaMin$ method.
\begin{small}
\begin{verbatim}
  1.   deltaMin(i)
  2.    j = xPy[i][getMin(x[i])]
  3.    setMax(y[j],yPx[j][i])
  4.    FOR (k = xlb[i] to getMin(x[i])-1)
  5.     j = xPy[i][k] 
  6.     setMax(y[j],yPx[j][i]-1)
  7.    END FOR LOOP
  8.    xlb[i] = getMin(x[i]) 
\end{verbatim}
\end{small}
Lines 1 to 3 are as the original. The next four lines (lines 4-7) cycle through each of the values that have been 
removed from the head of $dom(x[i])$ since the last call to $deltaMin(i)$ (line 4). $y[j]$, which  the removed value
corresponds to, is then found (line 5), and then all values that are not strictly greater than her 
preference for $x[i]$ are removed from $dom(y[j])$ (line 6). The lower bound of the man variable 
$x[i]$ is then updated (line 8).

\subsubsection{No enhancements to deltaMax(j)}
We now consider the situation where some process, other than a proposal, removes values
from the tail of $dom(y[j])$, i.e. when the maximum value of $dom(y[j])$ changes.
The $deltaMax$ method will be called, and the instance continues to be stable as all values remaining in $dom(y[j])$ 
corresponding to men $w_j$ prefers to the removed values. However, we need to prevent bigamy, by removing $w_j$ from 
the corresponding $dom(x)$ variables removed from the tail of $dom(y[j])$, and this is just what $deltaMax$ does. 
Therefore, no enhancement is required.

\subsubsection{Incomplete Lists (SMI)}
The encoding can also deal with incomplete preference lists, i.e. instances of the stable marriage problems
with incomplete lists (SMI). For a SM instance of size $n$ we introduce the value $n+1$. The value $n+1$ must
appear in the preference lists $mpl[i]$ and $wpl[j]$ as a {\em punctuation} mark, such that any
people after $n+1$ are considered unacceptable. For example, if we had an instance of size 3 and a preference 
list $PL(m_{i})$ = (3,2) we would construct $mpl[i] = (3,2,4,1)$ and this would result in
the inverse $mPw[i] = (4,2,1,3)$. Consequently $x[i]$ would always prefer to be unmatched (assigned
the value 4) than to be married to $y[1]$. We now need to modify the $init$ method such that it 
sets the maximum value in $dom(x[i])$ to be $mPw[i][n+1]$. These modifications will only work in the full
implementation (i.e. it requires the above enhancements).

\subsubsection{Reversible integers}
In this encoding we have used two variable arrays which contain dynamic data. $yub$ and $xlb$ are initialised to 
$n$ and 1 respectively, but these values will be updated as the problem is being made arc-consistent. If we are only 
looking for the first solution then we need only use normal integers to hold these values. However, when the 
constraint solver backtracks and values that had been removed from the domain of a variable are reintroduced then 
the values held in $yub$ and $xlb$ will no longer be correct. To fix this problem we have to tell the solver that 
when it backtracks it needs to reverse the changes to $yub$ and $xlb$ as well as the variables domains. This is 
done by using a reversible integer variable. This class should be supplied in the constraint solver toolkit. The 
solver will then store the values of each of the reversible variables at each choice point and restore them on
backtracking.

\section{Computational Experience}
We implemented our encodings using the JSolver toolkit \cite{JSolver}, i.e. the Java version of ILOG Solver. 
In a previous paper \cite{sara05} we presented a specialised binary constraint (SM2) for the 
stable marriage problem, and presented some results comparing the SM2 constraint with the two constraint encoding
in \cite{cp01}. 
Here we show a chopped down version of those results, with the results obtained by running SM2N on the same 
set of test data included. The other model shown in the results table is the optimal boolean encoding (Bool) as
presented in \cite{cp01}. Our experiments were run on a Pentium 4 2.8Ghz processor with 512 
Mbytes of random access memory, running Microsoft Windows XP Professional and Java2 SDK 1.4.2.6 with an increased
heap size of 512 Mbytes. 

\begin{table}[htb]
\begin{center}
\begin{tabular}{|c|c|c|c|c|c|c|c|} \hline
 & \multicolumn{6}{c}{size $n$} \hfill \vline \\ \hline
model   & 100   & 200   & 400   & 600   & 800   & 1000  \\ \hline 
Bool    & 1.2   & 4.4   & ME    & ME    & ME    & ME    \\ \hline
SM2     & 0.23  & 0.5   & 1.82  & 4.21  & 8.02  & 12.47 \\ \hline 
SM2N    & 0.02  & 0.06  & 0.21  & 0.51  & 0.95  & 2.11  \\ \hline

\end{tabular}
\caption{Average computation times in seconds to produce the GS-lists, from 10 randomly generated 
stable marriage problems each of size $n$}
\label{tab1}
\end{center}
\end{table}

Our first experiment measures the time taken to generate a model of a given SM instance and make that 
model arc-consistent, i.e. to produce the GS-lists. Table \ref{tab1} shows the average time taken 
to produce the GS-lists for ten randomly generated instances of size 100 up to 1000. Time is measured in seconds, and 
an entry $ME$ means that an out of memory error occurred. We can see that the SM2N constraint dominates the
other models. 
\begin{table}[htb]
\begin{center}
\begin{tabular}{|c|c|c|c|c|c|c| } \hline
 & \multicolumn{6}{c}{size $n$} \hfill \vline \\ \hline 
model   & 100   & 200   & 400   & 600   & 800   & 1000   \\ \hline 
Bool    & 2.02  & 6.73  & ME    & ME    & ME    & ME     \\ \hline
SM2     & 0.47  & 1.97  & 10.13 & 27.27 & 54.98 & 124.68 \\ \hline 
SM2N    & 0.03  & 0.07  & 0.24  & 0.73  & 1.56  & 3.35   \\ \hline

\end{tabular}
\caption{Average computation times in seconds to find all solutions to 10 randomly generated stable 
marriage problems each of size $n$}
\label{tab2}
\end{center}
\end{table}

This second experiment measures the time taken to generate a model and find all possible stable matchings.  
Table \ref{tab2} shows the average time taken to find all solutions on the
same randomly generated instances used in the first experiment. Again it can be seen that the SM2N model
dominates the other models. In summary, when the boolean encoding solves a problem the n-ary constraint 
does so nearly 100 times faster, and the n-ary constraint can model significantly larger problems than the 
boolean encoding.

Tables 1 and 2 raise the following question, if the Bool encoding is optimal then why is it dominated by the 
SM2 encoding, when SM2 is $O(n^3)$ time and the Bool encoding is $O(n^2)$ time?
The main reason for this is that there is no significant difference in the
space required to represent variables with significant differences in domain size, because domains are
represented as intervals when values are consecutive.  Considering only the variables,
the Bool encoding uses $O(n^2)$ space whereas the SM2 model uses $O(n)$ space. For example, with $n = 1300$ the Bool
encoding runs out of memory just by creating the $2.1300^2$ variables whereas the SM2 model takes 
less than 0.25 seconds to generate the required 2600 variables each with a domain of 1 to 1300. 
Theoretically the space complexity of the constraints used by SM2 and Bool are
the same. In practise this is not the case as SM2 requires exactly $n^2$ constraints to solve a problem of size $n$
whereas Bool requires $2n + 6n^2$ constraints. Therefore the Bool encoding requires more variables and more
constraints, resulting in a prohibitively large model. The same argument also applies to the performance of the 
SM2N constraint, i.e. the n-ary constraint is more space efficient that the Bool encoding, is of the 
same time complexity, and this results in superior performance. The space and time complexities of
these models are tabulated below. Note that the $O(n^2)$ constraint-space for SM2N is a consequence of the
storage of the preference lists and their inverses.
\begin{table}[htb]
\begin{center}
\begin{tabular}{|c|c|c|c|} \hline 
                  & Bool     & SM2      & SM2N      \\ \hline 
time              & $O(n^2)$ & $O(n^3)$ & $O(n^2)$ \\ \hline 
constraints space & $O(n^2)$ & $O(n^2)$ & $O(n^2)$ \\ \hline
variables space   & $O(n^2)$ & $O(n)$   & $O(n)$   \\ \hline
\end{tabular}
\end{center}
\caption{Summary of the complexities of the three SM constraint models}
\label{tabComplexity}
\end{table}

\begin{table}[htb]
\begin{center}
\begin{tabular}{|c|c|c|c|c|c|c| } \hline
 & \multicolumn{6}{c}{size $n$} \hfill \vline \\ \hline 
problem   & 1000   & 1200   & 1400   & 1600   & 1800   & 2000   \\ \hline 
AC        & 2.11   & 3.12   & 5.93   & 8.71   & 11.59  & 20.19  \\ \hline
All       & 3.35   & 5.09   & 8.8    & 12.92  & 18.96  & 26.81  \\ \hline 
\end{tabular}
\caption{Average computation times in seconds from 100 randomly generated stable 
marriage problems each of size $n$}
\label{tab3}
\end{center}
\end{table}

This Third experiment shows how SM2N can handle larger problems. Table \ref{tab3} shows the average time taken
to both produce the GS-Lists and find all solutions for one hundred randomly generated instances of size 1000
up to 2000, again the times are in seconds.

\section{Sex equal optimisation}
The sex equal stable marriage problem (SESMP) as posed in \cite{gi89} as an open problem, 
is essentially an optimisation
problem. A male optimal solution to an SMP is where all men get there best possible choices from all possible 
stable matchings (and all women get there worst), and in a woman optimal solution all women are matched to there
best possible choices (and all men to there worst). A sex equal matching is where both the men and the women are 
equally well matched. This problem has been proven to be NP-Hard \cite{se93}.

In a $SESMP$ all men will have a score for each woman and all women will have a score for each man, man $m_i$'s 
score for woman $w_j$ is $mScore[i][j]$ and woman $w_j$'s score for man $m_i$ is $wScore[j][i]$. In an
unweighted $SESMP$ all scores will be the same as the preferences, so $mScore[i][j]$ would equal $mPw[i][j]$ and
$wScore[j][i]$ would equal $wPm[j][i]$. In a weighted $SESMP$ this is not so, but the same ordering must be maintained
meaning $mScore[i][j] < mScore[i][k]$ iff $mPw[i][j] < mPw[i][k]$. For any matching $M$ all men and women will 
score the matching determined by which partner they are match to in $M$. If man $m_i$ is matched to woman $w_j$
in matching $M$ then $m_i$ will give that matching a score of $mScore[i][j]$ and woman $w_j$ will give it a score of
$wScore[j][i]$. The sum of all scores given by men for a matching $M$ equals $sumM(M)$ and the sum of the women's 
scores 
is $sumW(M)$. A matching $M$ for an instance $I$ of the stable marriage problem is sex equal iff there exists no 
matching 
$M'$ such that the absolute difference between the $sumM(M')$ and $sumW(M')$ is less than the absolute difference 
between 
$sumM(M)$ and $sumW(M)$.

Because the values in the domains of the $x$ and $y$ variables are preferences, it makes finding an unweighted 
sex equal matching with $SM2N$ simple. All that is required is to add a search goal to minimise the absolute
difference between the sum of all $x$ variables and the sum of all $y$ variables.
We tested this using the same test data as in Table \ref{tab3} and the results are tabulated below.
These results can be compared to those in Figure 6 of \cite{sara05}, where the Bool encoding failed to
model problems with 300 or more men and women, and at $n = 1000$ the SM2 model was 
more than 15 times slower than the SM2N model.
We believe that this demonstrates the versatility of our constraint, in that we can easily use the constraint
as part of a richer problem.
\begin{table}[htb]
\begin{center}
\begin{tabular}{|c|c|c|c|c|c|c| } \hline
 & \multicolumn{6}{c}{size $n$} \hfill \vline \\ \hline 
problem   & 1000   & 1200   & 1400   & 1600   & 1800   & 2000   \\ \hline 
SE        & 3.65   & 5.02   & 8.73   & 14.44  & 17.59  & 22.44  \\ \hline
\end{tabular}
\caption{Average computation times in seconds to find all solutions to 100 randomly generated sex-equal stable 
marriage problems, each of size $n$, modelled using the SM2N constraint.}
\label{tab4}
\end{center}
\end{table}

\section{Implementation}
The SM2N constraint was originally developed using the choco constraints tool
kit, and the way the constraint has been introduced reflects that. In choco 
to implement a user defined constraint, the $abstractLargeIntConstraint$ class 
is extended. This class contains the methods $awake$, $awakeOnInf$,
$awakeOnSup$, $awakeOnRem$ and $awakeOnInst$. These methods are the equivalent 
of the ones used to introduce the constraint. $awake$ is the same as $init$, 
$awakeOnInf$ and $awakeOnSup$ are the same as $deltaMin$ and $deltaMax$ and
$awakeOnInst$ is the same as $inst$. To implement a constraint in Ilog JSolver 
we first state when the constraint needs to be propagated, i.e. when a domain value 
is removed, when the range changes (meaning the upper or lower bound changes) 
or just when a variable is instantiated. We then need to define a method that 
will handle propagation when such an event occurs. For the SM2N constraint we 
stated it was to be propagated every time the range of a variable changed. We
then used conditional statements to ascertain which bound had changed, and 
used the methods as presented above to handle the propagation.

\section{Conclusion}
We have presented a specialised n-ary constraint for the stable marriage problem, possibly with incomplete lists. 
The constraint can be used when 
stable marriage is just a part of a larger, richer problem. Our experience has shown that this constraint can be 
implemented in a variety of constraint programming toolkits, such as JSolver, JChoco, and Koalog. The complexity 
of the constraint is $O(n^2)$. Although this is theoretically equal to the optimal $O(n^2)$ complexity of the 
Boolean encoding in \cite{cp01}, our constraint is more practical, typically 
being able to solve larger problems faster. For example, we have been able to enumerate all solutions to instances 
of size 2000 in seconds, whereas in \cite{ecai02} the largest problems investigated were of size 60. We have also 
presented the first study of SESMP using a constraint solution, i.e. where the stable matching
constraints are part of a richer problem.

\section*{Acknowledgements}
We are grateful to ILOG SA for providing us with the JSolver toolkit via an Academic Grant licence. We would also 
like to thank our four reviewers.
\bibliographystyle{plain}

%
\end{document}